\documentclass[runningheads]{llncs}

\usepackage{comment}
\usepackage{booktabs}
\usepackage{float}
\usepackage{graphicx}
\usepackage{bbding}
\usepackage[hyphens]{url}
\usepackage{graphicx}
\begin{document}

\title{Investors Embrace Gender Diversity, Not Female CEOs: The Role of Gender in Startup Fundraising
}





%
\titlerunning{Investors Embrace Gender Diversity, Not Female CEOs}
%

\author{}
\institute{}

\author{Christopher Cassion \and
Yuhang Qian \and
Constant Bossou \and \\ Margareta Ackerman$^{\textrm{\Envelope}}$}
\authorrunning{C. Cassion et al.}
%
\institute{Santa Clara University, Santa Clara CA 95050, USA \\
\email{mackerman@scu.edu}}
\maketitle              
\begin{abstract}
The allocation of venture capital is one of the primary factors determining who takes products to market, which startups succeed or fail, and as such who gets to participate in the shaping of our collective economy. While gender diversity contributes to startup success, most funding is allocated to male-only entrepreneurial teams. In the wake of COVID-19, 2020 is seeing a notable decline in funding to female and mixed-gender teams, giving raise to an urgent need to study and correct the longstanding gender bias in startup funding allocation. 




We conduct an in-depth data analysis of over 48,000 companies on Crunchbase, comparing funding allocation based on the gender composition of founding teams. Detailed findings across diverse industries and geographies are presented. Further, we construct machine learning models to predict whether startups will reach an equity round, revealing the surprising finding that the CEO's gender is \emph{the} primary determining factor for attaining funding. Policy implications for this pressing issue are discussed. 




\end{abstract}

As gender equality continues to make strides across a wide range of industries from STEM to medicine, there is a critical sphere where bias persists: Compared to their male counterparts, women have little access to startup funds, restricting them from engaging in our economy at this critical level. According to Pitchbook, in 2019, female founders raised just 2.7\% of the total venture capital funding invested and mixed gender founding teams received 12.9\%~\cite{pitchbook2020vcfemale}. 

The economic impact of the COVID-19 pandemic is having severe consequences for female entrepreneurs. Compared with 2019, the first quarter of 2020 saw a decline in the proportion of deals made with female and mixed-gender teams and funding allocated to female teams. In the third quarter of 2020, funding given to female-only teams dropped to 1.8\% with mixed-gender teams receiving just 11.1\% ~\cite{pitchbook2020vcfemale}.  There is an urgent need for understanding the nature of this persistent bias and uncovering effective solutions for systemic change.



In the United States, only 10-15\% of startups are founded by women~\cite{ewens2020early}. Yet, the number of women starting companies is not the primary issue, the far more important problem is their lack of access to capital~\cite{forbes2019gender}. The funding gap between male and female founders is particularly high at the early stage of a venture, with an analysis of  California and Massachusetts startups revealing that female-led ventures are 63\% less likely than male-led ones to obtain VC funding~\cite{guzman2019gender}.



While it is generally well-known in the venture community that men have an easier time raising funds, much remains unclear. In order to ascertain effective solutions, it is necessary to gain insight into the nature of the problem. For instance, does having a woman on a founding team increase or decrease fundraising outcomes? What role does the gender of the CEO play compared to the gender of other founders? If funding is successfully raised, how does gender impact the amount raised? How much does gender matter in different geographic regions and across industries?

We perform an in-depth data analysis of over 48,000 companies on Crunchbase. Our analysis suggests the presence of bias against women across geographies and industries, which extends not only to female-only but also to mixed-gender teams with female CEOs. We also construct machine-learning models (Decision Tree, Random Forest, Logistic Regression, Gradient Boosted Trees, and Multi-layer Perceptron (MLP)) to predict whether a founding team will reach a priced funding round.\footnote{Raising a priced round is a major milestone that offers startups the means to  succeed.} Our findings show the CEO's\footnote{In startups, the role of CEO is most often taken by one of the founders. This is nearly ubiquitous at early stages.} gender to be the most important founder characteristic for predicting fundraising success, beating critical features including whether the founders attended top universities and the number of prior exits. We discuss the implications of these findings to the utilization of machine learning models in venture capital allocation, and make recommendations for systemic change.

\section{Background}

Gender plays a key role across the lifetime of an entrepreneurial journey: Women are less likely to become entrepreneurs than men~\cite{ruef2003structure} and less likely to get external funding once a new venture is founded \cite{yang2014s}. The funding gap between male and female founders is higher at the early stage of the venture than at later stages~\cite{coleman2010sources}. Women are 65\% less likely to get funded at early stages and 35\% less likely to be funded at later stages, when strong signals of growth are available~\cite{guzman2019gender}.




Consequently, women-owned businesses rely heavily on internal funding (ex. personal finances) rather than funding from others, both debt and equity, to finance their firms~\cite{coleman2010sources}. Even though the number of women-owned firms is increasing rapidly~\cite{greene2001patterns}, they are still left behind compared to their male counterparts in receiving external founding.



Previous work provides valuable insight into the role of gender in the allocation of Venture Capital funds. However, the data used in previous studies, such as those above, tends to be geographically limited (focusing on individual countries, often the US), or consisting of only  several hundred instances. Many questions remain to be answered: How wide is the gender gap in the allocation of venture capital funding across geographic regions and industries? Does gender diversity help or hinder fundraising outcomes? Does the gender of the CEO play a special role compared to other founders? 

In order gain a broader understanding into the nature and prevalence of gender bias in VC, we perform the most comprehensive analysis to date on the impact of gender on startup funding across geographies and industry verticals, utilizing both statistical methods and machine learning techniques. We are careful to account for the potential influence of the pipeline problem, whereby fewer women seeking to engage in entrepreneurship.\footnote{The pipeline problem is often perceived as the primary cause of the gender gap in startup funding allocation, suggesting that the gap would be eliminated if women were as interested as men in pursuing entrepreneurship. We devise and apply analysis methods that shed light into these issues in a manner that cannot be reduced to the pipeline problem.} The data analysis helps inform our policy recommendations, and we hope that it will support future research on resolving gender bias in startup funding allocation.


\section{Methodology}

We rely on Crunchbase data to attain a data set of over 48,000 companies along with founder information. We consider four gender compositions:
 founding teams consisting entirely of male  founders (male-only), founding teams consisting entirely of female founders (female-only), teams with at least one female and at least one male founder led by a male CEO (mixed male-led), and teams with at least one female and at least one male founder led by a female CEO (mixed female-led). Companies with these gender and leadership compositions are subsequently compared, with emphasis on funding raised across a variety of industries and geographies. We then construct machine learning models to ascertain the importance of the team's gender composition and the leader's gender in funding outcomes. 

\subsection{Data Collection}

The data was obtained from Crunchbase, which prides itself for being ``the leading destination for company insights from early-stage startups to the Fortune 1000.''\footnote{\url{https://www.crunchbase.com/}} Crunchbase provides two majors types of data: Information on companies and data on individuals in leadership positions. We separately retrieved both types of data as they include some non-overlapping features. For instance, gender information is only available as a founder attribute and is absent from the company description. 


Our final dataset is an integration of the company and founder data. We first downloaded data of 224,000 companies and 175,000 founders with attributes of interest to our analysis. We then combined the two datasets by matching the Company's Website attribute in both the founders and companies dataset to produce a new dataset of 63,462 data points. The combined dataset contains all the attributes of the companies and all the aggregated attributes of the founders dataset. We dropped all rows with missing values in the key attributed (headquarter region, total funding raised, and industry) and obtained a final dataset of 48,676 entries. 58.14\% of the companies are led by multi-member founding teams. It is worth noting that companies ranked higher by Crunchbase tend to have fewer missing values.

While not all founders are present in the  founders dataset, founders names are present in the company dataset as comma separated attributes. Whenever a founder's gender is missing from the founder data set, we rely on a machine-learning model for gender classification based on names.\footnote{We utilized the following name-based gender classifier: \url{https://github.com/clintval/gender-predictor}. We retrained the model, achieving an accuracy of 97.10\%.}

Another important aspect is identifying who is leading the startup. We define the leader as either the CEO, or  the sole founder for one person founding teams. In order to determine leadership, we inspect the job titles of all of the founders found on Crunchbase that have the company as their primary organization. We set if the company is male or female led based on the gender of the identified founder. Female-only and male-only companies are respectively female and male led.

We reclassified the industries attribute values by reducing over one hundred industries down to thirty by combining closely related industries, from amongst which twenty industries with over 300 companies each were selected. We then picked the first industry that each company provides as its primary industry. The company's headquarter region was used to identify its location. 



\subsection{Attribute Statistics}

\begin{figure}[!ht]
    \centering
    \includegraphics[scale = 0.3]{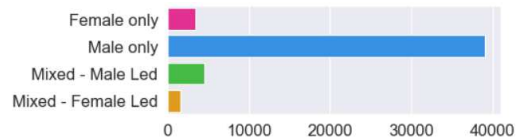}
    \caption{\textbf{Number of companies of each gender composition type}}
    \label{gender_number}
\end{figure}

Before delving into extensive analysis, we share some basic statistics about the data. As shown in Figure~\ref{gender_number}, overall founder gender distribution of the 48,676 companies in our dataset consists of 7.13\% female-only companies, 80.22\% male-only companies,  3.26\% mixed female-led companies, and 9.39\% mixed male-led companies (see Figure \ref{gender_number}) 

\begin{figure}[!htt]
    \centering
    \includegraphics[scale = 0.2]{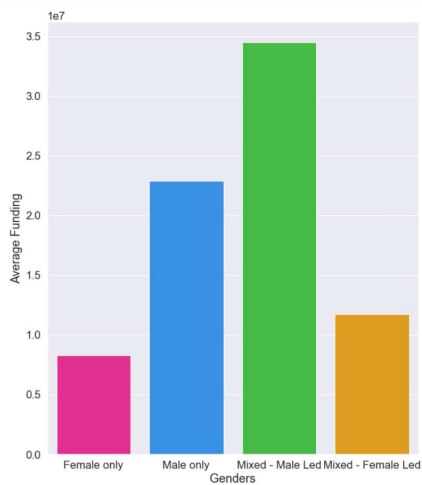}
    \caption{\textbf{ Average funding by  gender composition of founding teams}. Values in tens of millions of USD.}
    \label{gender_global_avg}
\end{figure}

Our analysis includes 20 industries, each consisting of at least 300 companies (See Figure~\ref{cate_number} for the list of industries). We omit locations with fewer than 1,500 companies, resulting in three major geographic regions, consisted of North America, Europe, and Asia-Pacific.
Since 64.01\% of companies are located in North America, we also include a detailed analysis focusing on companies based in the top four US startup hubs: Silicon Valley Bay Area, Greater New York Area, Greater Los Angeles Area, and Greater Boston Area. Lastly, 94.84\% of the startups in our data were founded on or after the year 2000. Please see the Appendix for additional information about the data set.

\section{Data Analysis}

In this section, we analyse the funding allocated to founding teams with different gender compositions. Results are reported across diverse industries and geographic regions.  




\subsection{Analysis by industry}

\begin{figure}[!ht]
    \centering
    \includegraphics[scale = 0.3]{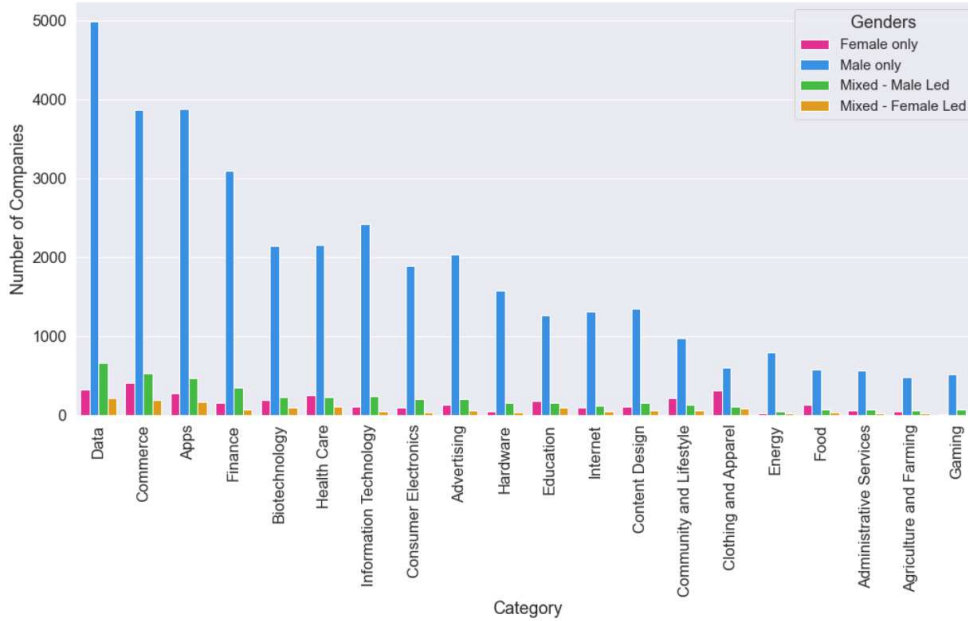}
    \caption{\textbf{Number of companies for each gender composition type by industry.}  All 20 industries are dominated by male-only teams.}
    \label{cate_number}
\end{figure}

We begin with an analysis of funding allocation by industry across the 20 most dominant industries identified in our data. As shown in Figure~\ref{cate_number}, there are far more male-only companies than female-only and mixed-gender companies  across all industries. The industries Data, Commerce and Apps have the largest number of companies while Gaming, Agriculture and Farming and Administrative Services have the fewest. In all but 5 out the 20 industries, the next biggest category is male-led, mixed-gender groups. 


\begin{figure}[!ht]
    \centering
    \includegraphics[scale = 0.3]{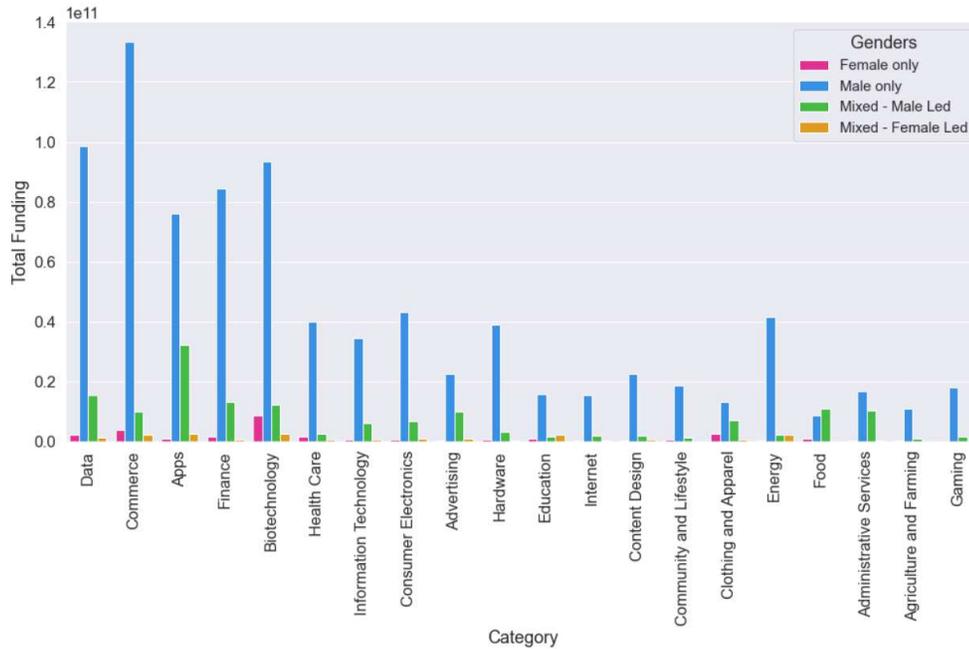}
    \caption{\textbf{Total funding allocation for founding teams with different gender composition across industries.} Values in in hundreds of billions of USD.  Total funding across all twenty one industries are dominated by companies founded by male-only founders except in the Food industry where mixed-gender male-led teams raised more funding than any other group type.}
    \label{cate_total_funding}
\end{figure}
\begin{figure}[!ht]
    \centering
    \includegraphics[scale = 0.3]{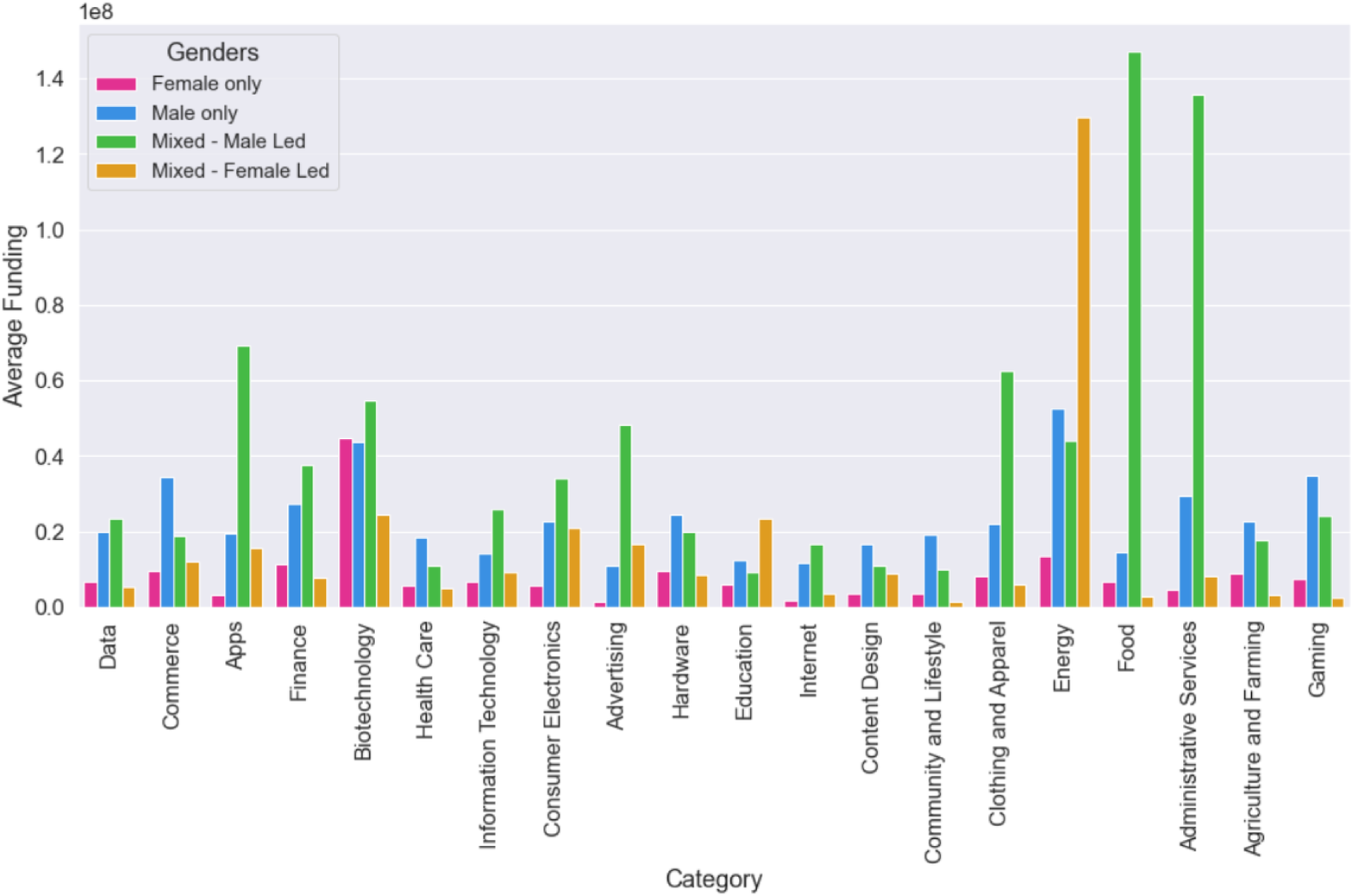}
    \caption{\textbf{Average funding allocation for founding teams with different gender composition across industries.} Values in hundreds of millions of USD. Of the 20 industries, average funding is highest for mixed-gender male-led teams in 11 industries and for male-only teams in 7 industries.}
    \label{cate_average_funding}
\end{figure}

As shown in Figure~\ref{cate_total_funding}, male-only and male-led mixed-gender teams receive the great majority of funding. In particular, male-only teams receive significantly 31 times more funding than female-only teams (statistic=11.0715, \emph{P}$<0.0001$). Male-only teams also get significantly 47 times more funding than mixed-gender female-led teams (statistic=5.8197, \emph{P}$<0.0001$). Similarly, mixed-gender male-led teams also receive significantly 8.5 times more funding than mixed-gender female-led (statistic=3.5987, \emph{P}$<0.0001$) and female only teams (statistic=4.2129, \emph{P}$<0.0001$). 



Male-only teams receive more funding in 19 of the 20 industries. In 18 of the 20 industries, there is a significant difference between the amount raised by male-only teams compared with female-only teams, the exceptions being agriculture \& farming and biotechnology, where the difference is not significant. On the other hand, the difference between male-only and male-led is often insignificant, with a significant difference found in only 4 of the 20 industries. 


For example, in the Data industry, male-only teams get significantly more funding than  mixed female-led teams (statistic=6.0181, \emph{P}$<0.0001$), and female only teams (statistic=4.6942, \emph{P}$<0.0001$), and insignificantly more than mixed male-led teams (statistic=0.6319, \emph{P}$=0.5276$). 
Similarly, in Commerce, male-only teams get significantly more funding than mixed female-led teams (statistic=3.4581, \emph{P}$=0.0005$) and female only teams (statistic=3.7047, \emph{P}$=0.0002$) and significantly more than mixed male-led teams (statistic=2.6360, \emph{P}$=0.0084$). Of the industries studies, Food is the only one where male-only teams did not raise the largest amount of total funding, however, the difference between male-only and mixed male-led teams was not significant (statistic=0.9567, \emph{P}$=0.3427$).






\subsection{Average funding by industry}


The pipeline problem, the fact that fewer women engage in entrepreneurship, is often perceived as the primary factor in the discrepancy in funding allocation. In order to gain insight into the nature of the issue beyond the pipeline problem, we consider the average funding allocated to teams that have successfully raised funds, comparing the amounts raised against the gender of the founding teams.  This analysis helps gain insight while offering an accessible demonstration of a potential gender gap to lay audiences.


As shown on Figure~\ref{cate_average_funding}, male-only founding teams and male led founding teams lead in average funding, receiving the highest amount of average funding across most industries (18 out 20). In 11 of the 20 industries, mixed-gender male-led team achieve the highest average funding, compared with 7 industries where male-only teams raise the most average funding. In industries including Food and Administrative Services  there is a substantial gap between average funding given to mixed male-led led teams and male-only, with the mixed teams raising a greater amount of funding. 

Of the twenty industries, there are only two industries (Energy and Education) where female-led teams receive more average funding. Notably, there are no industries where female-only teams raise the greatest amount of average funding. Unlike total funding, this persistent gap in average funding to startups that successfully raise funds cannot be explained by low numbers of women entrepreneurs. 


Comparing companies led by women, we find that in 9 of the 20 industries female-only teams receive more average funding than mixed-gender female-led startups.  



\subsection{Analysis by geography}
\begin{figure}[!ht]
    \centering
    \includegraphics[scale = 0.3]{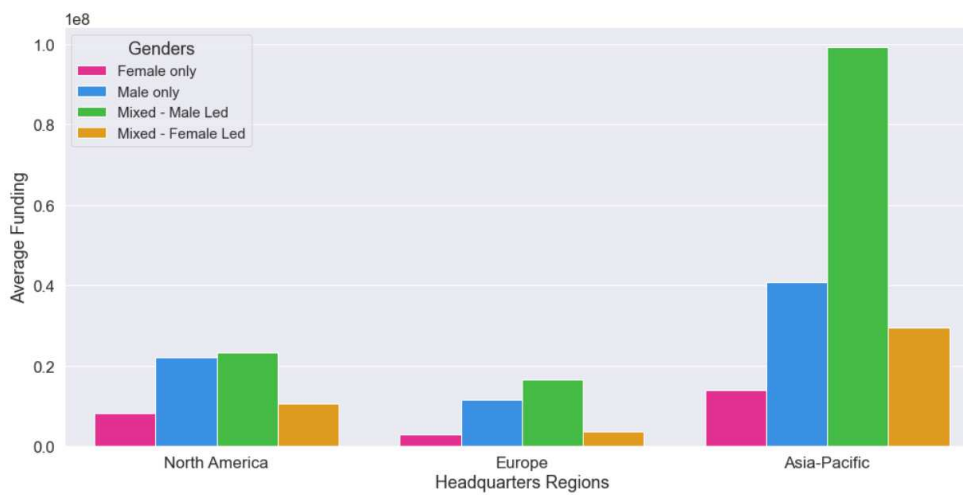}
     \caption{\textbf{Average funding allocation to founding teams with different gender composition across dominant continents.} Values in hundreds of millions of USD. All continents under consideration reveal the same raking, with male-led  mixed-gender teams  receiving  the  highest average funding, followed by male-only teams, then female-led mixed-gender teams, and finally female-only teams receiving the lowest amount of average funding.}
    \label{world_average_funding}
\end{figure}
\begin{figure}[!ht]
    \centering
    \includegraphics[scale = 0.3]{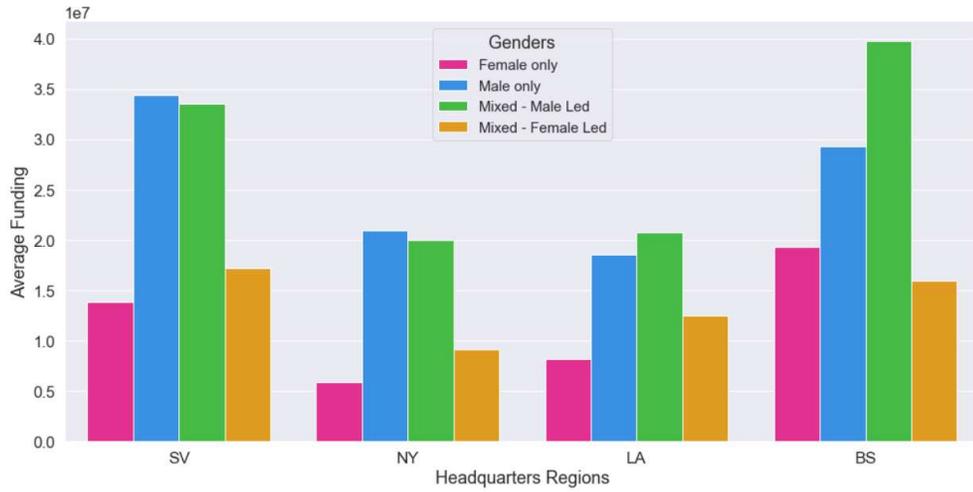}
    \caption{\textbf{Average funding allocation to founding teams with different gender compositions across US startup hubs.} Values in tens of millions of USD. In the US, Silicon Valley Bay Area and New York allocate the highest average funding to male-only teams, whereas male-led mixed-gender groups in Greater Los Angeles Area and Boston receive more average funding than all other group types.}
    \label{us_average_funding}
\end{figure}

\begin{figure}[!ht]
    \centering
    \includegraphics[scale = 0.3]{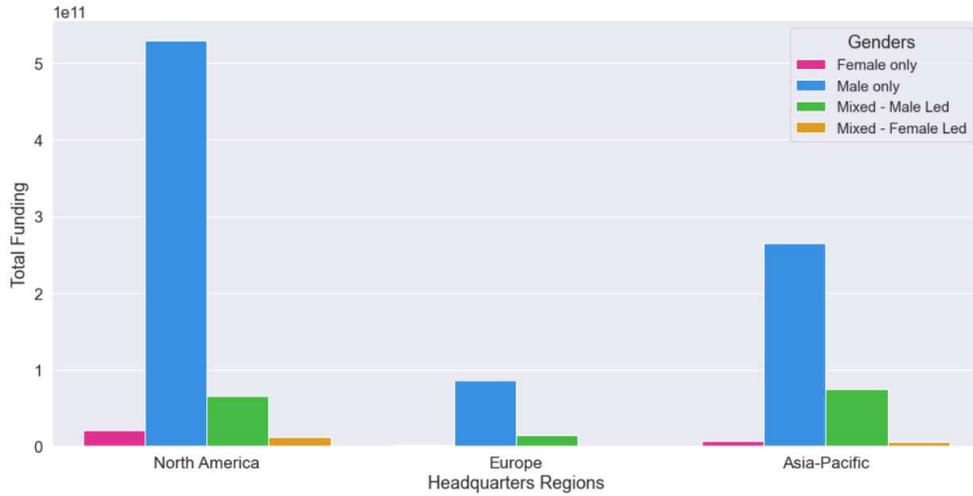}
    \caption{\textbf{Total funding by major geographic regions.} Values in hundreds of millions of USD. Male-only founding teams receive the greater amount of funding in all regions considered, while female-only teams and female-led teams receive the least founding.}
    \label{world_total_funding}
\end{figure}
\begin{figure}[!ht]
    \centering
    \includegraphics[scale = 0.3]{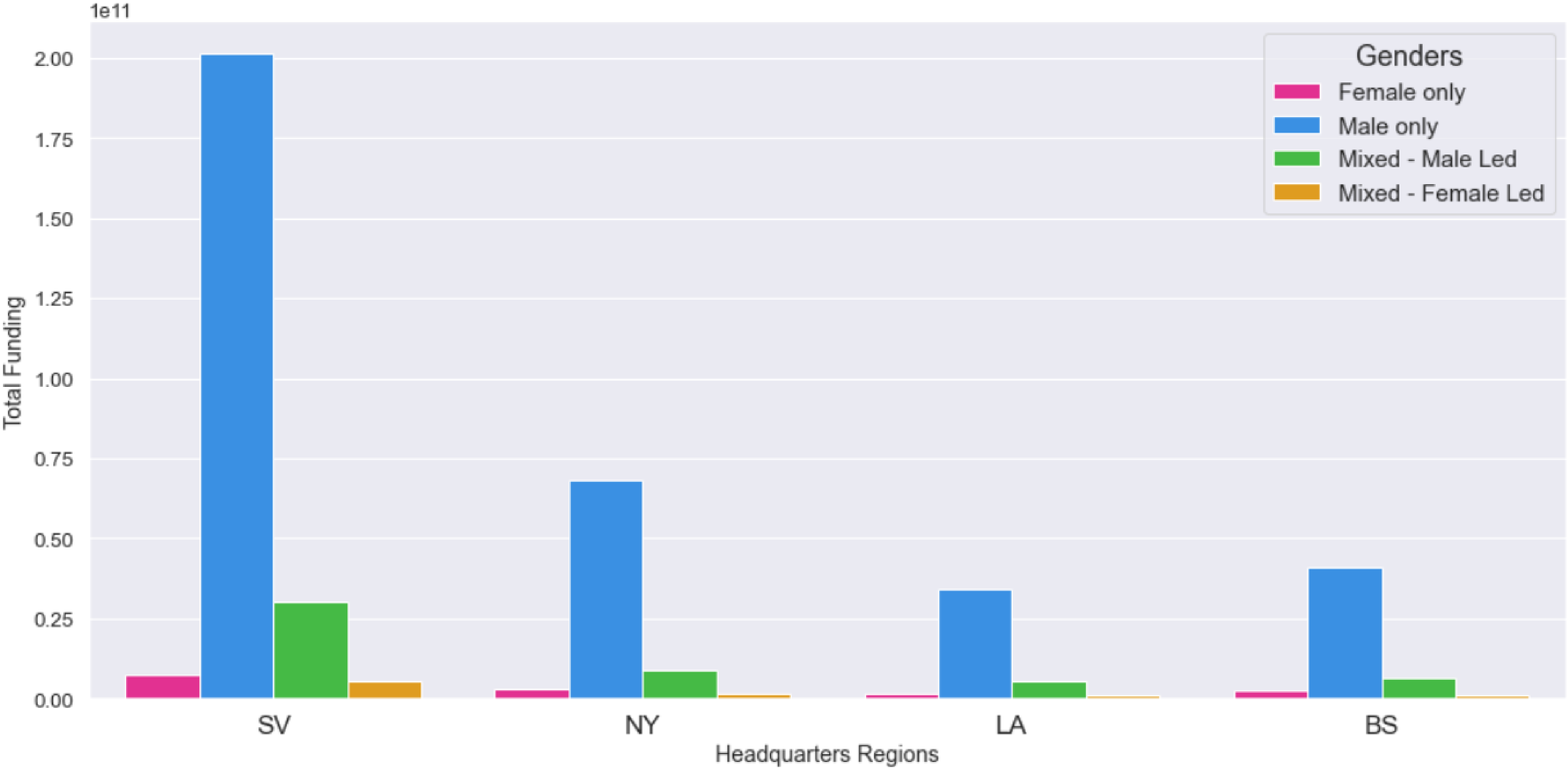}
    \caption{\textbf{Total funding by major US startup hubs.} Values in hundreds of billions of USD. Male-only founding teams receive the greater amount of funding in all regions considered, while female-only and female-led teams receive the least.}
    \label{us_total_funding}
\end{figure}



Considering average funding allocation, shown in Figure~\ref{world_average_funding}, we see the same raking by average funding allocation across all continents, with female-only teams receiving the lowest average funding, followed by female-led mixed-gender teams, then male-only teams, and finally mixed-gender male-led teams receiving the highest amount of average capital.

Analyzing startup hubs in the United States, shown in \ref{us_average_funding}, we discover that in Silicon Valley and New York male-only teams receive the highest average funding, narrowing beating mixed-gender male-led teams. LA and Boston follow the global trend of giving mixed-gender male-led teams the highest amounts of average funding, followed by male-only teams. In Silicon Valley, New York, and LA, female-only teams receive the least amount of average funding, followed by mixed female-led teams. However, in Boston, female-only teams receive more average funding than mixed-gender female-led teams.

In summary, companies with male CEOs receive greater funding across all continents and US startup hubs compared with companies with female CEOs. Mixed-gender teams perform well with respect to fundraising, often better than male-only teams, when they are led by male-CEOs.


When comparing total funding for different gender composition teams across continents (see Figure~\ref{world_total_funding}), we find that mixed-gender teams receive the great majority of funding in the three continents considered. Europe appears to be exhibiting the greatest preference for male-only teams, where such groups receive significantly over 65.5 times more funding than female-only companies (statistic=6.5061, \emph{P}$<0.0001$). European male-only teams raise significantly 92.9 more money than mixed female-led teams (statistic=7.6873, \emph{P}$<0.0001$). Comparing mixed gender teams, those led by men raise 15.3 more funding than female-led groups (statistic=2.0306, \emph{P}$=0.0427$). Male-only companies in Europe raise non-significantly 6 times more total funding than mixed-gender teams led by men (statistic=0.6232, \emph{P}$=0.5327$). 

In Asia-Pacific, male-only companies receive significantly over 37 times more funding than female-only companies (statistic=5.1950, \emph{P}$<0.0001$). Male-only teams raise insignificantly over 43 times more than mixed-gender female led teams (statistic=1.0063, \emph{P}$=0.3157$). Comparing mixed gender teams, those led by men raise insignificantly 12 times more funding than female-led groups (statistic=1.8761, \emph{P}$=0.0611$). Finally, male only teams raise 3.5 time more than male-led mixed-gender teams, not statistically significant (statistic=1.6322, \emph{P}$=0.1032$). 


When looking at total funding for different gender composition teams across US startup hubs (see Figure~\ref{us_total_funding}), similar to the continental analysis, male only teams receive the great majority of funding for the four major hubs. Silicon Valley exhibits some of the greatest preference for male founders, with male-only companies receiving significantly over 27 times more funding than female-only companies (statistic=2.9081, \emph{P}$<0.0038$) and significantly over 35 times more than mixed-gender female-led companies (statistic=2.7921, \emph{P}$<0.0054$).

In Los Angeles Area, male-only companies significantly receive over 23 times more funding than female-only companies (statistic=3.3092, \emph{P}$<0.0010$). Comparing male-only to mixed female-led companies, male-only raised insignificantly over 30 times more (statistic=1.0888, \emph{P}$=0.2787$). New York gives male-only companies over 21 times more funding than female-only companies \\
(statistic=6.8847, \emph{P}$<0.0001$) and over 37 times more than mixed-gender female-led companies (statistic=3.4533, \emph{P}$=0.0006$), both results being significant.

Finally, analysis of Boston area shows that  male-only companies receive about 16 times more funding than female-only companies (statistic=1.8629, \emph{P}$<0.0639$) and about 41 times more than mixed-gender female-led companies (statistic=1.4243, \emph{P}$<0.1599$), however here the results were not significant.

\section{Predictive Models}

Venture capitalists’ primary aim is to identify startups that will become successful in the future. As such, machine learning models have been playing an increasingly important role in the venture capital space (see, for example, \cite{krishna2016predicting}, \cite{halabi2010model} and ~\cite{arroyo2019assessment}).\footnote{Further, many venture capital firms built their own custom models which they do not make public in order to maintain a competitive advantage.} While predictive models can be used at any stage of investment, the problem is particularly challenging for early stage startups, prior to the availability of qualitative data on company performance. Prediction for later stage startups benefit from information on factors such as revenue and growth, making prediction significantly more accurate. On the other hand, early stage investments, which are often pre-revenue and precede product market fit, rely primarily on founder characteristics. 

One of primary risks with the utilization of machine learning models from an ethical perspective is the perpetuation and even amplification of existing biases. For instance, in the context of credit markets, Black and Hispanic borrowers are disproportionately less likely to gain
from the introduction of machine learning \cite{fuster_2017}. 



How much gender bias is present in startup data? To what degree does the utilization of machine learning models stands to perpetuate, or even amplify, gender bias in venture capital? We explore this direction by creating several machine learning models based on founder characteristics, the dominant characteristics available for early stage investments. We then analyze feature importance to ascertain how much the predictions rely on the gender composition of founding teams and the gender of the CEOs. Note that our exploration differs significantly from prior work in predictive modeling for startup success, since we are interested specifically in the importance of gender for attaining a priced funding round. By contrast, most work in the field aims to predict startup success by incorporating information about the startup itself, including quantitative success indicators such as total funding raised and number of employees. 



\subsection{Feature Selection}

In order to ascertain investor behaviour prior to having clear success indicators available, we focus exclusively on founder characteristics\footnote{As mention above, while a gender gap exits at all startup stages, investors are most reluctant to invest in women in the early stages, where female ventures are 65\% less likely to receive funding~\cite{guzman2019gender}}. However, it is essential to avoid including features that would be heavily altered by the target variable. For instance, social media presence stands to alter for founders who successfully raised funding. Similarly, information regarding investments made by the founders is heavily influenced by their entrepreneurial success, and are as such also omitted.



\subsubsection{Training Features}

To build the model, we first extract a set of features related to the founders from the aggregated dataset discussed in Methodology. The following features have been selected:
\begin{itemize}
    \item \textbf{Male Led}: Boolean variable that is True if the CEO or sole founder is Male, False otherwise
    \item \textbf{Gender Composition}: If the founding team is male-only, female-only, mixed male-led, or mixed female-led
    \item \textbf{Total Previously Founded Organizations}: Total number of companies previously founded by members of the founding team
    \item \textbf{Average Previously Founded Organizations}: The average number of companies previously founded by the founders 
    \item \textbf{Has Previously Founded Organizations}: Boolean variable indicating True if any of the founders previously founded an organization
    
    \item \textbf{Total Number of Exits}: Total number of exit events in which the founders participated 
    \item \textbf{Average Number of Exits}: Average number of exit events for the company founders
    \item \textbf{Has Exits}: Boolean variable indicating True if any of the founders had previously founded a company that had an exit event 
    
    \item \textbf{Total Number of Founders}: Number of founders of the company
    \item \textbf{Multiple Founders}: Boolean variable set to True if the founding team consists of two or more founders
    
    \item \textbf{Same Alma Mater}: Boolean variable indicating True if all of the founders went to the same university, False otherwise
    \item \textbf{\% from Top School}: Percentage of founders that went to a top 100 school \cite{top_universities_2020}
    \item \textbf{Top School Attended}: Boolean variable set to True if any of the founders went to a top 100 school\cite{top_universities_2020}

\end{itemize}


\subsubsection{Target Feature}
The goal of these experiments is to determine if a startup reached an equity funding round based on its founders. An equity round is when a startup sells shares of the startup in exchange for a large investment (generally well over a million). Equity rounds are important to the life cycle of startups largely because they provide a significant monetary influx into the company and represents a vote of confidence from the Venture Capital community, which helps with subsequent rounds. 


We separate the dataset into two funding stage groups, pre-equity rounds and post-equity rounds. We define pre-equity rounds as those whose latest funding stage is an Angel Round, Pre-Seed, Seed Round, or Convertible Note. We define post-equity rounds as those that whose latest funding stage is Series A, Series B or beyond, or Corporate Rounds. Using that, we construct models to predict whether a founding team has reached a priced round.

\subsection{Model Analysis}
Using hyperparameter grid-search to obtain the best model of each type, we constructed the following models: (1) Decision Tree (DT), (2) Random Forest, (3) Logistic Regression (LR), (4) Gradient Boosted Trees (GBT), and (5) Multi-layer Perceptron (MLP), for each of the worldwide and US data. MLP had the highest accuracy on worldwide data, at $63.73\%$. Similar results were found for US data with MLP giving the highest accuracy of $63.61\%$. Figure \ref{WorldModelsResults} and Figure \ref{USModelsResults} summarize the results for worldwide and US data, respectively. All models performed comparably, with worldwide accuracies varying by only 0.86\%. 


\begin{table}[!ht]
    \centering
    \begin{tabular}{|c|c|c|c|c|} 
     \hline
     Model & AUC & Precision & Recall & Accuracy \\  
     \hline\hline
     Decision Tree & 53.30 & 59.90 & 53.23 & 62.91\\ 
     \hline
     Random Forest & 53.20 & 59.78 & 53.17 & 62.86 \\
     \hline
     Logistic Regression & 52.70 & 61.38 & 52.73 & 63.00 \\
     \hline
     Gradient Boosted Trees & 53.00 & 60.04 & 53.04 & 62.88 \\
     \hline
     Multi-Layer Percepton & \textbf{53.80} & \textbf{63.92} & \textbf{53.80} & \textbf{63.72} \\ 
     \hline
    \end{tabular}
     \vspace{2mm}
    \caption{\textbf{Predictive model results for worldwide data.}}
    \label{WorldModelsResults}
\end{table}

Early stage predictions are known to be highly challenging. It is essential to emphasize that no information about the companies has been provided beyond founder features, in order to ascertain the impact of gender on early stage investing. It is unlikely that much higher accuracy is possible without incorporating features beyond the scope of founder characteristics.

\begin{table}[!ht]
    \centering
    \begin{tabular}{|c|c|c|c|c|} 
     \hline
     Model & AUC & Precision & Recall & Accuracy \\  
     \hline\hline
     Decision Tree & 54.20 & 58.36 & 54.17 & 60.68\\ 
     \hline
     Random Forest & 54.70 & 58.90 & 54.70 & 61.00 \\
     \hline
     Logistic Regression & 55.80 & 60.97 & 55.83 & 62.08 \\
     \hline
     Gradient Boosted Trees & 55.00 & 60.15 & 55.00 & 61.52 \\
     \hline
     Multi-Layer Percepton & \textbf{57.60} & \textbf{63.63} & \textbf{57.59} & \textbf{63.61} \\ 
     \hline
    \end{tabular}
     \vspace{2mm}
    \caption{\textbf{Predictive model results for the US dataset.}}
    \label{USModelResults}
\end{table}

\subsubsection{Feature importance in tree based models}
Considering feature importance enables us to ascertain how significant are gender-related characteristics compared with other founder attributes, such as prior exits or whether founders attended top schools. Table \ref{USModelsResults} and Table \ref{DTRFFeatureImportance} detail the feature importance for the Decision Tree and Random Forest models.\footnote{We report feature importance for the interpretable tree-based models, emphasizing that all models obtained comparable accuracy. Importance analysis for other models are left for future work.} The results show that whether a company is led by a male CEO is by far the most important feature in the decision tree model, and also the top feature in the random forest model for both the worldwide and US-only datasets. 

 \begin{table}[!ht]
    \centering
    \begin{tabular}{|c|c|c|c|} 
     \hline
     Feature & Decision Tree & Feature & Random Forest \\  
     \hline\hline
     Male Led & 25.62$\%$ & Male Led & 15.50$\%$ \\ 
     \hline
     \% from top school & 18.81$\%$ & Total Number of Founders & 12.97$\%$ \\
     \hline
     Has Exits & 18.81$\%$ & \% from top school & 11.26$\%$ \\
     \hline
     Same Alma Mater & 10.10$\%$ & Same Alma Mater & 9.12$\%$ \\
     \hline
     Total Number of Founders & 6.99$\%$ & Has Exits & 7.86$\%$ \\ 
     \hline
    \end{tabular}
    \vspace{2mm}
    \caption{\textbf{Decision Tree and Random Forest feature importance for top 5 features for worldwide data.} Whether the company is led by a male CEO is the most important feature for both the decision tree and random forest models, more important than the features addressing the number of prior exits and founders' alma mater. }
    \label{DTRFFeatureImportance}
\end{table}

\begin{table}[!ht]
    \centering
    \begin{tabular}{|c|c|c|c|} 
     \hline
     Feature & Decision Tree & Feature & Random Forest \\  
     \hline\hline
     Male Led & 19.34$\%$ & Male Led & 18.40$\%$ \\ 
     \hline
     Number of Exits & 19.29$\%$ & Number of Founders & 13.78$\%$ \\
     \hline
     \% from top school & 14.81$\%$ & \% from top school & 9.38$\%$ \\
     \hline
     Same Alma Mater & 12.46$\%$ & Same Alma Mater & 7.91$\%$ \\
     \hline
     Number of Founders & 8.22$\%$ & Avg Number of Exits & 7.21$\%$ \\ 
     \hline
    \end{tabular}
     \vspace{2mm}
    \caption{\textbf{Predictive model feature importance on US data.} According to both the decision tree and random forest models, the most important feature for reaching a priced round in the US is whether the company consists of only male founders.}
    \label{USModelsResults}
\end{table}

For the Gradient Boosted Tree (GBT) model on US data, the top 5 features are ranked as follows: whether the founding team is male-only (14.46$\%$), number of founders (13.04$\%$), percent of founders from top schools (12.03$\%$), number of exits (11.46$\%$), and if all the founders have graduated from the same university (10.07$\%$). For the worldwide GBT model, where GBT achieved the second lowest accuracy of the models created, the top feature is the number of founders. 


In summary, feature importance analysis for the tree-based models indicates that in most instances, for both worldwide and US-only datasets, gender is key to fundraising success. We find that the main indicator of whether a startup will reach a priced funding round centers on gender, either the gender of the CEO or whether the team consists entirely of male founders. 

\section{Conclusions and Recommendations}

Our analysis suggests the presence of a pervasive and substantial bias against female-led  startups across geographies and industries. 
Looking at average funding, where only startups that have raised funds are considered, lets us eliminate the pipeline problem as a a primary explanation for discrepancies in funding allocation. The analysis reveals that, across all but 2 of the 20 industries considered, male-led teams received the highest average funding. Across all three continents in our analysis, North America, Europe, and Asia-Pacific, the highest average funding went to mixed-gender teams led by male CEOs. In all three continents, the least average funding went to female-only teams, followed by female-led mixed-gender teams. 

Among US startup hubs, Silicon Valley and New York gave the highest average funding to male-only teams, while LA and Boston gave greatest average support to mixed-gender teams led by male CEOs. As in the continental analysis, female-only teams receive the lowest average funding, following by mixed-gender teams with female CEOS. 


Our ML-based analysis reveals gender characteristics to be of highest importance amongst founder features for reaching a prices round, in particular, more important that traditionally prized characteristics pertaining to whether the founders have attended top universities or had prior exits. Worldwide analysis reveals the gender of the CEO to the most important feature. On US data, machine learning modelling shows that the most important characteristic tends to be whether all founders are male. 

In summary, we find that across all geographic regions and the great majority of industries, companies a male CEO have much better funding outcomes those with female CEO. The gender of the CEO appears to be \emph{the most} important factor in fundraising. With no exceptions across geographies and industries, our results show that startups led by male CEOs raise more money than startups raised by female CEOs, irrespective of the gender of the rest of the founding team. On the other hand, having women as founders but not CEOs improves funding results in some (but not all) cases, sometimes by a substantial margin. Across all geographies (but not all industries), female-led companies achieve better funding outcomes when they include a male co-founder. 




\subsection{Implications for machine learning modeling for investment decisions}

Our machine-learning analysis reveals that CEO's gender to be the most important amongst founder characteristics for attaining a priced funding round. In particular, gender composition was found to be more important than characteristics that are known to be prized by venture capitalists, such as the number of prior exits or the founder's alma mater. This surprising finding not only reveals the primary role of gender in venture capital allocation, but also warns of potential pitfalls when applying machine learning models to investment decisions. 

Machine learning models in other spheres, such as credit markets, have already been shown to inflate biases~\cite{fuster_2017}. We recommend exercising caution when building machine learning models for startup success prediction, in order to reduce the impact of gender on the resultant decision making. Most importantly, features directly capturing the gender of the founders should be omitted.\footnote{Gender information has been incorporated into previous ML models for startup success, see, for example~\cite{arroyo2019assessment}.} 

Further, we observe that features such as prior exits, while not directly capturing bias, may play an important role in perpetuating it. With longstanding low access to startup funding, women have much lower chances of having had previous exits. 

\subsection{Discussion and policy implications}

One of the most important findings of this analysis is the critical role of a CEO's gender for fundraising outcomes, even in mixed-gender teams. This is notable because in startups, particularly at early stages, division of labour amongst founder is a less clear cut than in mature companies. Thus, it unlikely that a mixed-gender company's performances in terms of investor returns will be impacted on the basis of whether a male or female co-founder is designated as the CEO. 

Yet, fundraising is often handled by the CEO, making the founder identified as such the primary link between the startup and any potential investors. 
Even when other founders are present, the CEO is expected to lead the discussion on behalf of their startup. Consequently, any bias against women, implicit or otherwise, is likely to manifest most strongly if the CEO is female. 


The critical role of the CEO in funding outcomes may thus be reduced to differences in how investors treat men and women. Prior research points to desperate treatment of men and women during startup pitches. In a study on interactions at TechCrunch Disrupt in New York City, investors asked men to expand on how the plan to reach success, whereas women were asked to defend themselves against failure \cite{kanze2018we}, which hindered women's ability to raise funds. Notably, both male and female investors exhibited this bias against female founders.\footnote{There is a prevalent notion that the key to eradicating gender bias in  startup investing lies in increasing the number of female investors. This view is oversimplified and potentially misleading. Both men and women are highly prone to bias against women~\cite{united2020tackling}. While increasing gender diversity amongst investors is important for a variety of reasons, tackling gender bias against female founders calls for more comprehensive solutions.} Further research is needed to elucidate the impact of implicit gender bias on fundraising outcomes and uncover the reasons behind ubiquitous lower propensity towards investing in female CEOs across the globe.


Our findings show that across all continents considered and some US startups hub mixed-gender are given higher average funding, when they are male-led. This likely stems from the inherent advantage of mixed gender teams. Gender balanced teams perform better than male-dominated teams in terms of sales and profits \cite{hoogendoorn2013impact}, and gender diverse executives teams are 21\% more likely to yield higher financial returns \cite{hunt2018delivering}. The venture capital firm First Round reported that their investments in companies with at least one female founder were meaningfully outperforming their investments in all-male teams~\cite{firstround}. In fact, their investment in companies with a female founder performed $63\%$ better than their investments with all-male founding teams~\cite{firstround}.

The benefit of a gender diversity helps mixed-gender teams raise funding, but only when they are  led by a male CEO.  The consistently lower funding allocation, both in total and on average, to mixed-gender teams when they are led by women CEOs points to the severity of gender bias in startup capital allocation. 

 In recent years, a number of Venture Capital firms emerged with the mandate to invest in teams with at least one female founder. However, these investment firms tend to be (1) late stage, and (2) utilize a ``follow'' investment strategy, investing only after another VC firm makes a substantial investment and sets the deal terms. This does little to help increase the number of female founders. 
 
Our findings suggest that importance of investing in female-led companies to correct the gender bias in capital allocation. We recommend the formation of venture capital firms with the mandate to invest in companies with female CEOs. Investing in women-led, mixed-gender teams should allow investors to benefit from the performance boost of gender diversity, while helping to correct the long standing bias against female business leaders. Investors  with expertise to lead early stage deals applying such practiced can further reap the benefit of early investing, receiving large equity in promising deals.

\bibliographystyle{unsrt}
\bibliography{references}

\section{Appendix}
This appendix includes additional information on the data used in our analysis. 



\begin{table}[H]
\centering
\caption{Total Number and Percentage of Companies per Region}
\begin{tabular}{lrr}
Geography & \# of companies & Percentage \\ 
North America & 30,212 & 62.07\% \\
Europe & 8,932 & 18.35\% \\
Asia-Pacific & 7,968 & 16.37\% \\
Latin America & 1,308 & 2.69\% \\
Gulf Cooperation Council & 256 & 0.53\% \\
Total & 48,676 & 100.0\% \\
\end{tabular}
\label{table3}
\end{table}   

\begin{table}[H]
\centering
\caption{Total Number and Percentage of Companies per Region}
\begin{tabular}{lrr}
Geography & \# of companies & Percentage \\ 
Silicon Valley Bay Area & 7,611 & 46.98\% \\
Greater New York Area  & 4,420 & 27.28\% \\
 Greater Los Angeles Area & 2,416 & 14.91\% \\
Greater Boston & 1,754 & 10.83\% \\
Total & 16,201 & 100.0\% \\
\end{tabular}
\label{table4}
\end{table}   

\begin{table}[H]
\centering
\caption{Total and Percentage of Companies per Industry}
\begin{tabular}{lrr}
Verticals & \# of companies & Percentage \\ 
Data & 6,200 & 12.74\% \\
Commerce & 5,008 & 10.29\% \\
Apps & 4,789 & 9.84\% \\
Finance & 3,679 & 7.56\% \\
Information Technology & 2,822 & 5.80\% \\
Health Care & 2,755 & 5.66\% \\
Biotechnology & 2,663 & 5.47\% \\
Advertising & 2,442 & 5.02\% \\
Consumer Electronics & 2,236 & 4.59\% \\
Hardware & 1,826 & 3.75\% \\
Education & 1,690 & 3.47\% \\
Content Design & 1,670 & 3.43\% \\
Internet & 1,575 & 3.24\% \\
Community and Lifestyle & 1,388 & 2.85\% \\
Clothing and Apparel & 1,107 & 2.27\% \\
Energy & 884 & 1.82\% \\
Food & 828 & 1.70\% \\
Administrative Services & 729 & 1.50\% \\
Gaming & 613 & 1.26\% \\
Agriculture and Farming & 612 & 1.26\% \\
Others & 3,160 & 6.49\% \\
Total & 48,676  & 100.0\% \\
\end{tabular}
\label{table2}
\end{table}

\end{document}